\documentstyle[12pt]{article}
\begin{document}
\title{Velocity and density profiles of granular flow in channels
using lattice gas automaton} \author{Gongwen Peng\thanks{email:
peng@phys.ocha.ac.jp} and Takao Ohta\thanks {email:
ohta@phys.ocha.ac.jp}\\ \small Department of Physics, 
Ochanomizu University, Tokyo 112, Japan} \maketitle
\begin{abstract}
We have performed two--dimensional lattice--gas--automaton simulations
of granular flow between two parallel planes. We find that the
velocity profiles have non--parabolic distributions while
simultaneously the density profiles are non--uniform. Under non--slip
boundary conditions, deviation of velocity profiles from the parabolic
form of newtonian fluids is found to be characterized solely by ratio
of maximal velocity at the center to the average velocity, though the
ratio depends on the model parameters in a complex manner. We also
find that the maximal velocity ($u_{max}$) at the center is a linear
function of the driving force ($g$) as $u_{max} = \alpha g - \delta$
with non--zero $\delta$ in contrast with newtonian fluids. Regarding
density profiles, we observe that densities near the boundaries are
higher than those in the center. The width of higher densities (above
the average density) relative to the channel width is a decreasing
function of a variable which scales with the driving force ($g$),
energy dissipation parameter ($\epsilon$) and the width of the system
($L$) as $g^{\mu} L^{\nu}/\epsilon$ with exponents $\mu = 1.4 \pm 0.1$
and $\nu = 0.5 \pm 0.1$. 
A phenomenological theory based on a scaling argument is presented
to interpret these findings.
\end{abstract}
\vspace{0cm}
\newpage

\noindent
{\bf I. Introduction}\\
\indent

Granular materials exihibt many interesting
phenomena. The segregation of particles
\cite{Williams,Haff,Rosato,Devillard}, heap formation and convection
cells under vibration \cite{Farady,Evesque,Taguchi,Gallas}, and
anomalous sound propagation \cite{Liu} are just a few examples. Such
phenomena occur because the dynamical responses of granular media are
quite complex and different from those of usual solids, fluids and
gases \cite{Jaeger,Jaeger2,Hayakawa,Hansen,Mehta}. Considering the
complexity in the dynamics, one is tempted to study first the granular
materials in simple geometries and proceed then to more complicated
situations. However, even in the simplest geometries such as hoppers
and tubes, their flow under uniform external driving force (such as
gravity) still shows complex dynamics
\cite{nature,Baxter,Poschel,Ristow,Jysoo,Peng,Matsushita}. For tubes,
one can observe density waves in granualr flow when the width of the
tubes is narrow enough. Here the friction of the wall, as well as the
dissipation among the grains themselve, plays an important role in the
occurence of density waves. \\ 

The purpose of this paper is to study
granular flow in channels (different from tubes with wider width) and
compare it with flow of Newtonian fluids. Comparison between granular
flow and ordinary fluids is useful because it links with the knowledge
that we have gained before. Besides the obvious contrast between the
discretness of granular materials and continuum of fluids, a key
difference is that collisions of particles in granular flow are
inelastic (energy--dissipative).  In fluids, channel flow is a typical
and fundamental topics in rheology (Poiseuille flow). Under the
non-slip boundary conditions valid for viscous fluids, Newtonian
fluids have a parabolic velocity profile across the channel.  To claim
that granular flow is non-Newtonian flow, a non--parabolic velocity
profile is a clear and fundamental evidence. In this paper, we will
report the non--parabolic shape of velocity profiles and
non--uniformly distributed density profiles in granular channel
flow. \\

Study on granular materials has a long history in engineering. Much
engineering literature is devoted to understanding how to deal with
these materials. However, it is only after the pioneering work of
Bagnold \cite{Bagnold} that theoretical work has begun to
progress. Instead of focusing on the detailed dynamics of individual
particles, granular hydrodynamics \cite{Haff1,Jenkins} treats the
granular media like ``fluids''. One defines a set of macroscopic
quantities like particle density, velocity, and granular
temperature. By assuming local equilibrium \cite{Li} one can write
down the equations based on mass conservation, momentum conservation
and energy balance. These equations are similar to the Navier--Stokes
equations for fluids. \\

Using the idea of hydrodynamics, Savage
\cite{Savage} has studied the channel flow of cohesionless granular
materials and obtained theoretically the velocity profiles. He
obtained the velocity profiles under non--slip boundary conditions for
different parameter $R$ (which depends on the system parameters and on
the density at the channel center). In the limiting case of $R=0$
(which corresponds to the situation where density are uniformly
distributed), the non--dimensional velocity profile has the form of $v
= 1 - x^{3/2}$ where $v$ is the velocity normalized to 1 with respect
to the maximal velocity at the center, x is the distance from the
channel center divided by the half--width of the channel. For larger
$R$, the velocity profiles become more blunted (but without simple
analytical form). Savage \cite{Savage} performed also experiments to
measure the velocity profile and found that his experiments
corresponded to the case of $R=0.3$. Since he did not intend to
compare his results with those for fluids, his theory did not recover
the parabolic velocity profile in any limiting case as he included
from the begining that the viscosity is proportional to the shear rate
\cite{Bagnold}. However, we find from his experimental data that the
velocity is very close to have a parabolic shape. \\

In the following we
will report our numerical results on the velocity profiles. We find
that when the dissipation is very weak the velocity profiles are close
to the parabolic shape while in the other cases they are more blunted
than the parabolic form. As mentioned above, we realize that the key
difference between fluids and granular flow is that the latter is
energy dissipative. Therefore, when the parameter controlling the
energy dissipation is switched off, the theory should recover that of
fluids. Furthermore, we observe that density near the boundaries are
higher than those at the channel center. Our numerical findings are
consistent with the earlier experimental observations by Soo $et$ $al$
\cite{Soo}: ``the nature of the concentration, mass flow, and velocity
distributions of solid particles is such that the concentration
increases toward the wall of the pipe, mass flow decreases toward the
wall, and velocity is less than or equal to that of the stream at the
core''. \\

In this paper we study the channel flow by computer simulations. Since
a general theory for granular media is not yet available, people have
used various computational methods to get better understanding about
the complicated rheological behavior of granular media. Among the
different methods are molecular dynamics (MD)
\cite{Haff,Ristow,shearflow,LeeHans}, Monte Carlo simulations
\cite{Devillard,MC}, the diffusing void model \cite{Hong}, event
driven algorithms \cite{luding} and cellular automaton \cite{CA}.  So
far the most widely used method is MD \cite{MDbook} which simulates
the granular materials on a ``microscopic'' level (the grain's level). 
MD has been recognized to be very successful in simulating granular
materials \cite{Hans}.  MD needs, however, much computer time to give
reasonable results. To calculate the velocity and density profiles one
needs long time average in order to get reasonable statistics. \\

The
same situation was also faced in classical fluid mechanics some years
ago when Frisch, Hasslacher and Pomeau \cite{FHP} proposed
lattice--gas automata (LGA) as a novel alternative to the direct
solution of the equation of motion. As a sort of primitive
molecular--dynamics system LGA offers the advantage of guaranteed
numerical stability coupled with extreme computational simplicity.
The basic idea behind LGA is that a properly defined cellular
automaton with appropriate conservation laws should lead to the
Navier--Stokes equations. A detailed comparison of simulational
results from LGA and the well--established theory of Newtonian fluids
in Poiseuille flow can be found in Refs. \cite{Kadanoff,BHS}. The main
check to the LGA for the Newtonian dynamics was the parabolic velocity
profile in channel flows \cite{Kadanoff,BHS,Rothman}. \\

In this paper we employ
the LGA model of Ref. \cite{Peng} to study the velocity and density
profiles in granular channel flow. The LGA of Ref. \cite{Peng} is an
extension of the LGA of usual fluids by including energy dissipation
among the particles into the model. As emphasized above, energy
dissipation is a major mechanism by which granular flow differs from
ordinary fluids. The LGA models for granular materials were sussessful
in simulating granualr flow \cite{Peng,Alonso,Kertesz,Vollmar}. \\

This
paper is organized as follows. We describe the model in some detail in 
Sect. II. The results obtained by simulations are presented in Sect. III. 
A phenomenological argument is made in Sect. IV to understand the numerical
results. 
A discussion is contained in Sec. V.  \\

\noindent
{\bf II. Simulational model}\\
\indent

We consider an LGA at integer time
steps $t=0,1,2,\cdots$ with $N$ particles located at the sites of a
two--dimensional triangular lattice. The arrangement of the triangle
lattice is depicted in Fig. 1 (a).  At each site there are seven
Boolean states which refer to the velocities, $\vec{c}_k
(k=0,1,2,\cdots 6)$. Here $\vec{c}_k (k=1,2,\cdots 6)$ are the nearest
neighboring (NN) lattice vectors and $\vec{c}_0=\vec{0}$ refers to the
rest (zero velocity) state. Each state can be either empty or occupied
by a single particle. Therefore, the number of particles per site has
a maximal value of $7$ and a minimal value of 0. In this paper we use
the number of particles per site as density which can therefore be
greater than 1.0. The time evolution of the LGA consists of a
collision step and a propagation step. In the collision step particles
change their velocities due to collisions and in the subsequent
propagation step particles move in the directions of their velocities
to the NN sites where they collide again.\\

The system is updated in parallel. Only the specified collisions shown
in Fig. 1 can deviate the trajectories of particles. All collisions
conserve mass and momentum.\\ 

For two-- and three--body collisions, we have the probabilistic rules
shown in Fig. 1 (b). The probablity that a configuration may take
place is shown next to the configuration. If the parameter $\epsilon$ is
nonzero, it means that energy can be dissipated in the collision. \\ 

Collisions with rest particles may produce more than one rest
particles on that site. This is allowed temporarily as in the Fig. 1
(c). However, immediately after the collision step, the extra rest
particles randomly hop to NN sites until they find a site with no rest
particle and there they stop. \\

We use no--slip boundary conditions at the channel walls 
which are parallel to the y--axis and periodic
boundary conditions along the channel. No--slip conditions are
employed by allowing that any particle colliding with the wall along
any of the three possible directions bounces back into the incoming
direction. \\

We incorporate the driving force, namely, gravity, in the way that
Kadanoff $et$ $al$ \cite{Kadanoff} have used. The direction of gravity
is along the y--axis, i.e., the downward in Fig. 1a.  After each time
step we randomly select a lattice site and, if possible, apply one of
the forcing rules: (i) a rest particle goes into motion with equal
probability along one of the two lattice directions which form an
angle of $30^0$ with the direction of gravity; (ii) a moving particle
change its velocity by a unit vector along the direction of gravity if
the resulting vector is possible on the triangle lattice used.  Each
successful application of a forcing rule adds one unit of momentum to
the system. The forcing process is repeated until the desired amount
of momentum (we label it as $G$) has been transfered to the system;
fractional amounts of momentum to be added to the system are
accumulated across time steps until they sum to an amount greater than
1, at which time one additional unit of momentum is added to the
gas. The actual forcing scheme is slightly more complicated since it
must compensate for inhomogeneity in the momentum and number densities
due to the macrosopic flow (see also Ref. \cite{Kadanoff}). The
forcing algorithm randomly selects a lattice row and column and then
searches along that column until it finds a site where a forcing rule
may be successfully applied. This guarantees that forcing operations
will be uniformly distributed across the width of the channel, despite
variations in the mass and momentum densities. It is noted that the
magnitude of the gravity cannot be extremely large in our model system
due to the following fact: on one hand the maximal velocity at the
channel center increases with the increasing gravity, on the other
hand the LGA only guarantees a velocity less than unit. \\

\noindent
{\bf III. Simulational Results}\\
\indent

 We evolve the system according to
the collision rules defined above.  The initial configuration of the
system is set to be random in the sense that every state (except the
rest state) of each site is randomly occupied according to a
preassigned average density $\bar{\rho}$. We discard the
configurations in the first period of several thousand time steps
until we are sure that the system is in the steady state where energy
input by gravity is averagely equal to the energy dissipation in the
system. Fig. 2 shows a typical curve of the total kinetic energy
relaxation starting from a random configuration at time step
$t=0$. \\

After the kinetic energy curve becomes flat, we then make the
necessary averge over space and time as follows. Let us label the
particle number of state $k (k=0,1,2,\cdots,6)$ on the $i$th lattice
site with coordinates $(x_i, y_i)$ at time step $t$ as
$n_k(x_i,y_i,t)$. As depicted in Fig. 1(a), the x--coordinates of
lattice sites take values of integers and half--integers of $0, 1/2,
1, \cdots, L-1$, while the y--coordinates have values of $0, 1/2, 1,
\cdots, M-1$ with $M$ even. The total number of lattice sites is $N=
LM - M/2$. In this paper, we always keep $M = 2L$ and fix the average
density $\bar{\rho} = 1.5$ unless otherwise specified. \\

We are interested in the density profile $\rho(x)$ and velocity
profile in y--direction $u(x) = V_y(x)$ with x integer
\begin{eqnarray}
\rho(x) &=& \displaystyle \frac{1}{2TM} \sum_{t=t_0+1}^{t_0+T}\sum_{i=1}^{N}
\sum_{k=0}^{6}{n_k(x_i,y_i,t)\delta([x_i]-x)} \\[5mm]
u(x) &=& \displaystyle \frac{1}{2TM} \sum_{t=t_0+1}^{t_0+T}\sum_{i=1}^{N}
\sum_{k=0}^{6}{c_k^{(2)} n_k(x_i,y_i,t)\delta([x_i]-x)} 
\end{eqnarray}
where the function $[z]$ takes the integer value of its variable
$z$ and function $\delta(z)$ equals to one when z=0 and zero
otherwise. $\vec{c}_k = (c_k^{(1)}, c_k^{(2)}) = (cos(\pi(k-1)/3),
sin(\pi(k-1)/3))$ for $(k=1,2,\cdots,6)$ are the
unit vectors of the lattice bond orientation (and $\vec{c}_0 = \vec{0}$). 
We actually make 
average over two column whose x--coordinates have the same integer
part, they are columns with $x_i$ (integer) and $x_i + 1/2$ (half--integer),
therefore, the normalization factor is $\frac{1}{2TM}$ where T is the
time period over which we make average starting from time step
$t_0$. T is usually several tens of thousand time steps.\\

As a first step, we check that when the dissipation parameter
$\epsilon$ is set to zero we recover the parabolic velocity profile of
Newtonian fluids. This is shown in Fig. 3(a) for system size $L=64$
with average density $\bar{\rho}=1.8$. 
From Fig. 3(a) we can estimste the magnitude of the kinematic viscosity
equal to $0.35$ according to Ref. \cite{Kadanoff}.
Meanwhile, the density profile in
Fig. 3(b) is just a flat curve, showing that density is uniformly
distributed.\\

Fig. 4(a) and Fig. 4(b) show the velocity and density profiles when
energy dissipation with $\epsilon=0.01$ 
is present in the system. We see clearly that the
parabolic velocity profile is no longer kept in this case, but more
blunted. 
In fact, a parabolic curve demanding the
maximum value at the center equal to the numerial data and zero at the
two boundaries deviates substantially from the numerical data.
It is also remarked that the density is not uniformly distributed, in
contrast with those of Newtonian fluids. \\

Regarding the velocity
profile, we can generally write down the following equation:
\begin{equation}
u(x)=\displaystyle u_{max}(\epsilon, g, L)
\tilde{u}(\frac{x}{L}, \epsilon L^{\sigma_1},
 g L^{\sigma_2})
\end{equation}
where $u_{max}$ is the maximum velocity.
Here gravity is taken to be system size
independent quantity $g = G/N$ where $G$ is the total amount of
momentum added to the system after each time step and $N$ is the total
number of lattice sites.\\

A simple measure of the flatness of velocity profiles is the ratio
$\gamma = u_{max}/\bar{u}$ where $\bar{u}$ is the mean value of
$u(x)$. It is equal to $1.5$ for Poiseuille flow of Newtonian fluids
and $1.0$ for a perfectly flat profile.
Note that $\gamma$ can be written as
\begin{equation}
\label{gamma}
\gamma^{-1} = \int_0^1 \tilde{u}(x,\epsilon L^{\sigma_1},g L^{\sigma_2}) dx
\end{equation}

Fig. 5 displays the ratio $\gamma$ versus gravity for two different
dissipation parameters $\epsilon = 0.01$ and $\epsilon = 0.02$ with
different channel widths.  From Fig. 5 we see $\gamma$ depends on the
channel width, degress of dissipation and gravity. We find that $\gamma$ can be solely used to 
characterize the shape function $\tilde{u}(x,\epsilon L^{\sigma_1},g
L^{\sigma_2})$. Taking two set of different parametres which leads to
almost identical $\gamma$, e.g., $\{\epsilon_1 = 0.01, G_1=8.0, L_1 =
64\}$ and $\{\epsilon_2 = 0.02, G_2=10.0, L_2 = 48\}$ resulting
$\gamma_1=1.27594$ and $\gamma_2=1.28298$, we show in Fig. 6 that
their shape functions are almost identical. Note that both set of
parameters $\{\epsilon, g, L\}$ share no common value but their
resulting $\gamma$ are almost identical. 
For comparison, the velocity profile for a different value of
$\gamma=1.35906$ is also plotted in Fig. 6. As expressed in
Eq. (\ref{gamma}), $\gamma^{-1}$ is the $0$th moment of the
distribution function $\tilde{u}(x)$. Fig. 6 tells us that the
distribution function $\tilde{u}$ has such a property that as long as
the $0$th moment is determined higher order moments are determined. 
This may suggest that the parameters $\{\epsilon, g, L\}$ combine to
give a scalar on which $\tilde{u}(x)$ depends solely.\\

In order to fit the velocity data, we have tried the polynominal
fitting routines. We find that polynominal up to an order of $4$ is
not sufficient; most of the data can be fitted by a
polynominal up to an order of $6$:
\begin{equation}
\label{polynominal}
\tilde{u}(x) = v_0 - v_2 (x-c)^2 - v_4 (x-c)^4 - v_6 (x-c)^6
\end{equation}
where $x$ is the normalized coordinate in the interval [0:1] and $c$ is the coordinate of the channel center. The three velocity profiles are fitted using  Eq. (\ref{polynominal}) in Fig. 6. In the next section basing on a phenomenological theory we will propose a
formula  for velocity profiles where 
the dependences of these
fitting parameters on the model parameters $\{\epsilon, g, L\}$ are
clearly expressed and we will check our theory by the
simulational data.\\

For $u_{max}(\epsilon, g, L)$, we find that it is a linear function of
$g$ as $u_{max}(\epsilon, g, L) = \alpha g - \delta$ with non--zero
$\delta$. Fig. 7 displays four curves of $u_{max}$ for different
channel widths but with fixed $\epsilon$. 
The slope $\alpha$ depends on $L$ but not as $L^2$
which is valid for Newtonian fluids. 
If we increase $\epsilon$, the value of $\alpha$ decreases.
Non--zero $\delta$ means that the
driving force must exceed a non--zero threshold ($\delta/\alpha$) in
order to drive the granular media into motion.  From the statistics we
are not sure whether $\delta$ is independent of $L$ but one can see
from Fig. 7 that the depenence is very weak, if any. Futhermore, 
although not shown in Figure, we
find that $\delta$ have a tendency to approach zero as $\epsilon$
approaches zero, which recovers the case of Newtonian fluids. \\

Regarding the density profile, we observe that densities near the
walls are higher than those at the channel center, in contrast to the
uniform density distribution of Newtonian fluids.  A simple way to
characterize the spatial variation is to define a width over which
densities are higher than the average density. We plot such a width
($\Delta$) divided by channel width in Fig. 8 (a) against gravity $g$
for different system sizes and different dissipation
parameters. Generally, $\Delta/L$ decreases with increasing $g$ and
increasing $L$ but increases with increaing $\epsilon$. The scattered
data in Fig. 8(a) can be scaled to one curve as shown in Fig. 8 (b)
where we plot $\Delta/L$ versus $g^{\mu} L^{\nu}/\epsilon$. We find
that the exponents are $\mu = 1.4 \pm 0.1$ and $\nu = 0.5 \pm 0.1$.\\

We have checked that the above observations do not change their
properties qualitatively as we change the average density
$\bar{\rho}$. The scaling exponents $\mu$ and $\nu$ are independent of
the average density. \\

\noindent
{\bf IV. Phenomenological Theory}\\
\indent

In this section we present a phenomenological theory to interpret our numerical findings. We note that there are two important system parameters, namely, $g$ and $\epsilon$: $g$ is the velocity change per particle per time step and $\epsilon$ is energy dissipation per particle per time step. Therefore, using dimension analyse, we have \\
\begin{equation}
\label{1-1}
g = \frac{d V}{d t}, \hspace{1cm} \epsilon =  \frac{d E}{d t}.
\end{equation}
where $V$ and $E$ are characteristic velocity and kientic energy of a particle. Hereinafter we put the particle's mass to be unity. Since $V^2 \sim E$, $gV/\epsilon$ must be dimensionless. \\

We may construct a quantity which has a dimension of velocity by using the channel width $L$ as 
\begin{equation}
\label{1-2}
V = \sqrt{g L}
\end{equation}
The above argument implies that $P$ defined as 
\begin{equation}
\label{1-3}
P = \frac{g V}{\epsilon} = \frac{g^{3/2} L ^{1/2}}{\epsilon}
\end{equation}
must be a fundamental dimensionless quantity. Eq. (\ref{1-3}) is consistent with the simulations illustrated in Fig. 8b where $\Delta/L$ is a function of a variable which scales as $g^{\mu} L^{\nu}/\epsilon$ with exponents $\mu = 1.4 \pm 0.1$
and $\nu = 0.5 \pm 0.1$.\\

Note that when $\epsilon = 0$, the system has another dimensionless quantity which can be derived from the newtonian hydrodynamics equation
\begin{equation}
\label{1-4}
g + \nu \nabla^2 v = 0
\end{equation}
where $\nu = \mu/\rho$ is the kinematic viscosity. The two terms in Eq. (\ref{1-4}) have the same dimension, therefore, 
\begin{equation}
\label{1-5}
Q =\frac{g L^2}{\nu V} = \frac{g L^2}{\nu \sqrt{gL}} = 
\frac{g^{1/2} L^{3/2}}{\nu}
\end{equation}
is another dimensionless quantity. Note in Eq. (\ref{1-5}) we have used Eq. (\ref{1-2}). $Q$ is in fact the square root of the Reynolds number.\\

Now, we consider the maximum velocity $u_{max}$ at the channel center. 
Simulations show that $u_{max}$ is a linear function of $g$ 
\begin{equation}
\label{2-1}
u_{max} = \alpha g - \delta.
\end{equation}

We determine the forms of $\alpha$ and $\delta$. First of all, they do not depend on $g$ by definition. Using Eq. (\ref{1-2}), $\alpha$ must take the form
\begin{equation}
\label{2-2}
\alpha = \sqrt{\frac{L}{g}} \hat{\alpha}(P, Q)
\end{equation}
The form of $\hat{\alpha}$ can be determined by the following conditions: (i) $\alpha$ (instead of $\hat{\alpha}$) is independent of $g$; (ii) when $\epsilon \rightarrow 0$ ($ P  \rightarrow \infty$), $u_{max} = gL^2/\nu$. The simplest interpolation formula for $\hat{\alpha}$ is given by
\begin{equation}
\label{2-3}
\hat{\alpha} = \frac{1}{c_1 P^{-1/3} + c_2 Q^{-1}}
\end{equation}
where $c_1, c_2$ are dimensionless positive numerical constants.
In fact we have 
\begin{equation}
\label{2-4}
\begin{array}{rcl}
\alpha g & = & \displaystyle \sqrt{gL} \frac{1}
         {\displaystyle c_1\left(\frac{\epsilon}{g^{3/2}L^{1/2}}\right)^{1/3} +
	 c_2 \frac{\nu}{g^{1/2}L^{3/2}}}
\\
[15mm]
& = &\displaystyle \frac{g\sqrt{L}} {\displaystyle c_1\left(\frac{\epsilon}{L^{1/2}}\right)^{1/3} +
	 c_2 \frac{\nu}{L^{3/2}}}
\end{array}
\end{equation}
Therefore,
\begin{equation}
\label{2-5}
\alpha=  \displaystyle \frac{L} {c_1\left({\epsilon L}\right)^{1/3} +
	 c_2 \nu/L}
\end{equation}
When $\epsilon=0$, we recover the well--known result:
\begin{equation}
\label{2-6}
u_{max} \sim \frac{g L^2}{\nu}
\end{equation}
When $\epsilon \rightarrow \infty$, Eq. (\ref{2-5}) shows
\begin{equation}
\label{2-7}
\alpha = \frac{L^{2/3}}{c_1 \epsilon^{1/3}}
\end{equation}
Our numerical results also show that $\alpha$ decreases with increasing $\epsilon$, in the trend of Eq. (\ref{2-7}). The $L^{2/3}$ is even reached in the simulations with $\epsilon = 0.02$. This is illustrated in Fig. 9. From Eq. (\ref{2-5}) one may see that as long as $\epsilon$ is non--zero, the term with $c_1$ dominates the denominator for large $L$, leading to the validity of  
Eq. (\ref{2-7}).\\

Now we determine $\delta$. It should vanish at $\epsilon=0$ and be independent of $g$ for any non--zero value of $\epsilon$. Using again the fact of Eq. (\ref{1-2})  $\delta$ should be given by 
\begin{equation}
\label{2-9}
\delta = c_3 \sqrt{g L} P^{-1/3} f(P/Q^3) = c_3 (\epsilon L)^{1/3} f(P/Q^3)
\end{equation}
where $c_3$ is a positive constant. Note that $P/Q^3 = \nu^3/(\epsilon L^4)$ is independent of $g$. It seems that the form of $f(x)$ can not be specified phenomenologically. If we impose the fact that $\delta$ is insensitive to $L$ as observed in simulations, $f(x)$ must be $f(x) \sim x^{1/12}$ so that we have
\begin{equation}
\label{2-10}
\delta = c_3 (\epsilon L)^{1/3} \nu^{1/4} \epsilon^{-1/12} L^{-1/3} = c_3 (\epsilon \nu)^{1/4}
\end{equation}

Now we turn to the shape function $\tilde{u}(x)$ of velocity profile. Since from the simulations we know the velocity profile can be well fit using polynominal up to $6$th order, we propose phenomenologically
\begin{equation}
\label{vel_shape}
\tilde{u}(x) = 1 - 4 u_2 (x-1/2)^2 - 16 u_4 (x-1/2)^4 - 64 u_6 (x-1/2)^6
\end{equation}
where the coefficients  $u_2, u_4, u_6$ are functions of $\epsilon$, $g$ and $L$. \\

As expressed in Eq. ({\ref{gamma}}), the $0$th moment of $\tilde{u}(x)$ is characterized by $\gamma$. We find numerically that the scattered data of $\gamma$ in Fig. 5 can be collapsed to one curve by using a combination of the dimensionless quantities $P$ and $Q$: $Q^4/P = g^{1/2} L^{11/2} \epsilon / \nu^4$. Fig. 10 illustrates this point by plotting $\gamma$ against $\Gamma = 10^{-5}g^{1/2} L^{11/2} \epsilon$ for the data of Fig. 5. We propose that the  coefficients  $u_2, u_4, u_6$ in Eq. (\ref{vel_shape}) are functions of $\Gamma$ and we determine their forms as follows.\\

Setting $x=0$ in Eq. (\ref{vel_shape}) and requiring $\tilde{u}(0) = \tilde{u}(1) = 0$, we know that $u_2, u_4, u_6$ are related by
\begin{equation}
\label{vel_relation}
 u_2 + u_4 + u_6 = 1
\end{equation}
So we need only to determine $u_2$ and $u_4$ which must satisfy the following conditions: (i) when $\epsilon = 0$, $u_2 = 1$, $u_4 = 0$ which recovers the parabolic velocity profile for fluids; (ii) when  $\epsilon \rightarrow \infty$, the velocity profile becomes completely flat and therefore $u_2 = 0$, $u_4 = 0$.We propose $u_2$ and $u_4$ as functions of $\Gamma$:
\begin{equation}
\label{u_2}
u_2 = \frac{1}{1 + d \Gamma}
\end{equation}
\begin{equation}
\label{u_4}
u_4 = \frac{e \Gamma}{1 + h \Gamma^2}
\end{equation}
where $d$, $e$, $h$ are constants. Substituting the above expressions into Eq. (\ref{vel_shape}) and integrating that equation we obtain 
\begin{equation}
\label{integ}
\begin{array}{rcl}
\gamma^{-1} &=& \displaystyle \frac{90 - 20 u_2 - 6 u_4}{105}\\
[8mm]
            &=& \displaystyle \frac{90 - \frac{20}{1 + d \Gamma} - \frac{6 e \Gamma}{1 + h \Gamma^2}}{105}
\end{array}
\end{equation}
We ckeck the above expressions by fitting the numerical data of $\gamma$ using Eq. (\ref{integ}). This is shown  in Fig. 10. One sees that the fitting is very well except for $\gamma$ close to 1.0 (corresponding to rather flat velocity profiles where $(x-1/2)^8$ might be necessary in Eq. (\ref{vel_shape}).). \\  

Finally we make a remark that velocity profile such that $ \tilde{u}(x) = 1 - u_{\omega} (x-1/2)^{\omega} - 4 u_2 (x-1/2)^2$ with $0 < \omega < 2$ is not in accord with $\gamma$ shown in Fig. 5.\\     

\noindent
{\bf V. Discussions}\\ 
\indent

In this paper we have numerically studied the
granular flow in channels using a two--dimensional lattice gas
automaton model. We have explored the dependences of velocity profile
on dissipation, gravity and channel width and compared them with those
of Newtonian fluids. Parabolic velocity profile is no longer valid as
long as there is energy dissipation in the system. However, the
deviation from parabolic profile depends on the degree of energy
dissipation in a smooth way (there is no sharp transition near
$\epsilon = 0$). This may explain the fact that the experimental data
of Savage \cite{Savage} is close to a parabolic velocity profile,
though he did not notice this point in his paper. However, the
velocity profile can also be very flat depending on the dissipation
and the channel width. The observation of Soo \cite{Soo} confirms this
point experimentally. Our results also show that the velocity profiles
are more blunted than the parabolic form. We do not find any velocity
profile that is sharper than parabolic shape (i.e., our $\gamma$ are
all less than $1.5$), in contrast with Savage's theory \cite{Savage}
which predicted that profiles sharper than parabolic ones are possible
to exist. The maximal velocity at the center is a linear function of
gravity as $u_{max} = \alpha g - \delta$ but with non--zero
$\delta$. Here $\alpha$ behaves no longer as $L^2$ as for Newtonian
fluids. Density profiles are no longer uniform but are higher near the
boundaries than at the center. From the simulations we note that this
nonuniformity is caused by the rest particles while the moving
particles are distributed uniformly. This may have a link to the
clustering of dissipative systems \cite{Zanetti}. The boundaries which
are represented by rough walls seem to serve as clustering seeds due
to the non--slip boundary conditions we used.  An earlier experiment
by Soo \cite{Soo} also showed that density increases toward the wall,
as we find here.  It is interesting to note that the recent experiment
of Pouliquen and Gutfraind \cite{Pouliquen} showed that density near
the wall is lower than that at the center. We attribute this
contradiction with our numerical results to the different boundary
conditions. In their experiments, considerable slip velocities were
observed at the walls.  Both increasing and decreasing density
profiles toward the walls in different regimes were obtained
theoretically in Ref. \cite{Goodman} and both kinds of experimental
evidences were cited there.\\

In our model we find that the width of the region of 
higher densities (than the average density) relative to the channel width
is a decreasing function of a variable which scales with the driving
force ($g$), energy dissipation parameter ($\epsilon$) and the width
of the system ($L$) as $g^{\mu} L^{\nu}/\epsilon$ with exponents $\mu
= 1.4 \pm 0.1$ and $\nu = 0.5 \pm 0.1$. Such a scaling was not found
in any previous investigations.
We have shown that $P=g^{3/2} L^{1/2}/\epsilon$ is a fundametal
dimensionless parameter in the present system. Combined this with $Q = g^{1/2} L^{3/2}/ \nu$, all of the results obtained in simulations have been successfully understood.\\

In this paper we have used non--slip boundary conditions at the
channel walls. Unlike in viscous fluids, non--slip boundary conditions
are not necessarily satisfied in every case of granular flow
\cite{Ehrichs}. In fact, the general boundary conditions for granular
flow is still lacking. However, as noted by Savage \cite{Savage}, the
non--slip condition may be realized by providing wall roughness of
the same order as that of the particle surface 
constituting the granular material, as he
achieved in his experiments \cite{Savage}. \\

\noindent
{\bf Acknowledgments}\\ 
\indent

This work was supported by
Grant--in--Aid of Ministry of Education, Science and Culture of Japan.  
G. P. thanks the Japan Society for the
Promotion of Science for a postdoctoral fellowship.  \\

\newpage
\noindent
{\bf Figure Captions}\\ 
\indent

Figure 1: (a) Sketch of the two--dimensional
triangle lattice. Gravity is along y--direction. One complete hexagon
is drawn; (b) Probabilistic collision rules for two-- and three--body
collisions. Thin arrows represent particles and small circles stand
for rest particles. The number next to a configuration is the
probability that the configuration takes place; (c) Collision rules
for moving particles with a rest particle. Immediately after the
collision, more than one rest particle on a site will hop to the
nearest neighbouring sites randomly until they find a suitable site
with no rest particle already there. \\

Figure 2: Total kinetic energy $E$ versus time step by running the LGA
starting from a random initial configuration. The kinetic energy of a
moving particle is taken to be a unit. The plotted curve was obtained
with $\epsilon = 0.01$. The total momentum added to the system each
time is $G=7.0$, channel width $L=64$ and average density
$\bar{\rho}=1.5$. \\

Figure 3: (a) Velocity profile for the LGA with $\epsilon$ set to
zero. Here the total momentum added to the system each time is
$G=1.0$, channel width $L=64$ and average density
$\bar{\rho}=1.8$. The curve is the best (least--square) fit to the
numerical data using a polynomial fitting rountine up to second order;
(b) Density profile corresponding to (a).\\

Figure 4: (a) Velocity profile for the LGA with $\epsilon =
0.01$. Here the total momentum added to the system each time is
$G=7.0$, channel width $L=64$ and average density
$\bar{\rho}=1.5$. The upper curve is the best (least--square) fit to
the numerical data using a polynomial fitting rountine up to second
order and the lower curve is a parabolic curve by demanding the
maxmial value at the center equal to the numerial data and zero at the
two boundaries; (b) Density profile corresponding to (a).\\

Figure 5: The ratio $\gamma = u_{max}/\bar{u}$ versus gravity $g$ for
different channel widths $L= 32, 48, 64, 128$ (from top to bottom)
with $\epsilon = 0.01$ (data points connected by solid lines) and $
\epsilon = 0.02$ (data points connected by dashed lines).\\

Figure 6: The normalized shape function $\tilde{u}(x)$ as a function
of $x$ in the inteveral [0:1]. Two curves corresponding to
\{$\epsilon=0.02, L=48, G=10.0 (g=0.00434028)$\} ($\diamondsuit$) and
\{$\epsilon=0.01, L=64, G=8.0 (g=0.00195312)$\} ($+$) are almost
identical, though the two set parameters share no common
value. Another curve ($\Box$) corresponding to \{$\epsilon=0.02, L=48,
G=5.0 (g=0.00217014)$\} is shown for comparison. This curve shares two
parameters $\epsilon$ and $L$ 
with the curve ($\diamondsuit$) but their difference is
obvious. Lines are best fit to the
numerical data using a polynomial fitting rountine up to $6$th order as expressed in Eq. (\ref{polynominal}).
The fitting parameters  are $\{v_0 = 0.999664, v_2 =
1.25789, v_4 = -0.51635, v_6 = 53.8739, c = 0.493339\}$ ($\diamondsuit$),
 $\{v_0 = 0.998367, v_2 = 1.35365, v_4 = -1.83773, v_6 = 55.4401, c
= 0.494999\}$ ($+$) and $\{v_0 = 0.997691, v_2 =
1.81431, v_4 = 4.97792, v_6 = 20.8538, c = 0.493152\}$ ($\Box$).\\

Figure 7: $u_{max}$ versus gravity $g$ for four different channel
widths $L= 32, 48, 64, 128$ (from bottom to top)
and for $\epsilon=0.02$. 
The lines are
least--squares linear fit to the numerical data. Four lines cut the
vertical axis at non--zero positions which are very close
to each other.\\

Figure 8: (a) The width of higher densities (than average density)
relative to the channel width $\Delta/L$ are plotted versus the
gravity $g$ for different channel widths and different dissipation
parameters. A: \{$\epsilon=0.01, L=32$\}, B: \{$\epsilon=0.01,
L=128$\}, C: \{$\epsilon=0.02, L=48$\}, D: \{$\epsilon=0.02, L=64$\},
E: \{$\epsilon=0.02, L=128$\}, F: \{$\epsilon=0.04, L=48$\}, G:
\{$\epsilon=0.04, L=64$\}. (b) Data collapse of (a): $\Delta/L$ are
plotted versus $g^{\mu} L^{\nu}/\epsilon$ with $\mu = 1.4$ and $\nu =
0.5$.\\

Figure 9: $\alpha$ versus $L^{2/3}$ for $\epsilon = 0.01$ ($\diamondsuit$)
and $\epsilon = 0.02$ ($+$). \\

Figure 10: Data collapse of Fig. 5: $\gamma$ versus $\Gamma = 10^{-5} g^{1/2} L^{11/2} \epsilon$. Symbols are the same as those in Fig. 5. Solid line is the best fit to the numerical data using Eq. (\ref{integ}). The logarithm of horizontal axis is made for clarity of displaying data. The fitting parameters in Eq. (\ref{integ}) are $d = 0.0097748$, $e=-0.350605$ and $h=0.00268941$.

\newpage
\begin{thebibliography}{1}
\bibitem{Williams} J. C.~Williams, Powder Technol.{\bf 15}, 245
(1976).
\bibitem{Haff} P. K.~Haff and B. T.~Werner, Powder Technol.{\bf 48}, 239
(1986).
\bibitem{Rosato} A.~Rosato, K. J.~Strandburg, F.~Prinz, and R.
H.~Swendsen, Phys. Rev. Lett. {\bf 49}, 59
(1987).
\bibitem{Devillard} P.~Devillard, J. Phys. France {\bf 51}, 369
(1990).
\bibitem{Farady} M.~Faraday, Philos. Trans. R. Soc. London {\bf 52}, 299
(1831).
\bibitem{Evesque} P.~Evesque and J.~Rajchenbach, Phys. Rev. Lett. {\bf
62}, 44(1989).
\bibitem{Taguchi} Y. H.~Taguchi, Phys. Rev. Lett. {\bf
69}, 1367(1992).
\bibitem{Gallas} J. A. C.~Gallas, H. J.~Herrmann, and S.~Sokolowski,
Phys. Rev. Lett. {\bf 69}, 1371(1992).
\bibitem{Liu} C--h.~Liu and S. R.~Nagel,
Phys. Rev. Lett. {\bf 68}, 2301(1992).
\bibitem{Jaeger} H. M. Jaeger and S. R.~Nagel,
Science  {\bf 255}, 1523(1992).
\bibitem{Jaeger2} H. M. Jaeger and S. R.~Nagel, and R. P.~Behringer, 
Rev. Mod. Phys.  {\bf 68}, 1259(1996); H. M. Jaeger and S. R.~Nagel, 
and R. P. Behringer, Phys. Today  {\bf 49}, 32(1996).
\bibitem{Hayakawa} H.~Hayakawa, H.~Nishimori, S.~Sasa, and Y.-h.~Taguchi, 
Jpn. J. Appl. Phys. Part 1  {\bf 34}, 397(1995).
\bibitem{Hansen} {\em Disorder and Granular Media}. edited 
by A.~Hansen and D.~Bideau (North--Holland, Amsterdam, 1992).
\bibitem{Mehta} {\em Granular Matter}. edited by A.~Mehta 
(Springer, Heidelberg, 1994).
\bibitem{nature} K.~L. Schick and A. A. Verveen, Nature {\bf 251},
599(1974).
\bibitem{Baxter} G. W.~Baxter, R. P.~Behringer, T.~Fagert, and G.
A.~Johnson, Phys. Rev. Lett.  {\bf 62}, 2825(1989).
\bibitem{Poschel} T.~P\"oschel, J. Phys. I France,  {\bf 4}, 499(1992).
\bibitem{Ristow} G.~Ristow and H. J.~Herrmann, Phys. Rev. E {\bf
50}, R5(1994).
\bibitem{Jysoo} J.~Lee, Phys. Rev. E {\bf 49}, 281(1994).
\bibitem{Peng} G.~Peng and H. J.~Herrmann, Phys. Rev. E {\bf 49}, 
R1796(1994); $ibid.$ {\bf 51}, 1745(1995).
\bibitem{Matsushita} S.~Horikawa, A.~Nakahara, T.~Nakayama, 
and M.~Matsushita, J. Phys. Soc. Japan {\bf 64}, 1870(1995).
\bibitem{Bagnold} R. A.~Bagnold, Proc. Roy. Soc. A  {\bf 225}, 49(1954).
\bibitem{Haff1} P. K.~Haff, J. Fluid Mech. {\bf 134}, 401(1983).
\bibitem{Jenkins} J. T.~Jenkins and S. B.~Savage, J. Fluid Mech. 
{\bf 130}, 187(1983).
\bibitem{Li} There is still controversy over the validity of 
granular hydrodynamics. See, e.g., Y. Du, H. Li, and L. P. Kadanoff, 
Phys. Rev. Lett. {\bf 74}, 1268(1995).
\bibitem{Savage} S. B. Savage, J. Fluid Mech. {\bf 92}, 53(1979).
\bibitem{Soo} S. L. Soo, G. J. Trezek, R. C. Dimick, 
and G. F. Hohnstreiter, Ind. Engng Chem. Fund. {\bf 3}, 
98(1964). see also Ref. \cite{Goodman}. 
\bibitem{Goodman} M. A. Goodman and S. C. Cowin, J. Fluid Mech. 
{\bf 45}, 321(1971).
\bibitem{shearflow} C.~S. Campbell and C.~E. Brennen, J. Fluid Mech.
  {\bf 151}, 167(1985); P.~A. Thompson and G.~S. Grest, Phys. Rev.
  Lett. {\bf 67}, 1751(1991); D.~M. Hanes and D.~L. Inman, J. Fluid Mech.
  {\bf 150}, 357(1985); O.~R. Walton and R.~L. Braun, J. Rheol. {\bf
    30}, 949(1986).
\bibitem{LeeHans} J.~Lee and H.~J. Herrmann, J. Phys. A {\bf 26}, 373(1993).
\bibitem{MC} A. Rosato, K.~J. Strandburg, F. Prinz and R.~H. Swendsen,
  Phys. Rev. Lett. {\bf 58}, 1038(1987); A.~D. Rosato, Y. Lan and
  D.~T. Wang, Powder Technol. {\bf 66}, 149(1991).
\bibitem{Hong} H.~Caram and D. C.~Hong, Phys. Rev. Lett. 
{\bf 67}, 828(1991); M. Y~Choi, D. C.~Hong, and Y. W.~Kim, 
Phys. Rev. E {\bf 47}, 137(1993); J. K.~Rudra and D. C.~Hong, 
$ibid.$ {\bf 47}, R1459(1993).
\bibitem{luding} S. Luding, E. Cl\'ement, A. Blumen, J. Rajchenbach
  and J. Duran, Phys. Rev. E {\bf 49}, 1634(1994).
\bibitem{CA} G.~W. Baxter and R.~P. Behringer, Phys. Rev. A {\bf 42},
  1017(1990); Physica D {\bf 51}, 465(1991).
\bibitem{MDbook} M.~P. Allen and D.~J. Tildesley, {\em Computer
    Simulations of Liquids}, Clarendon Press,
Oxford (1987).
\bibitem{Hans} H.~J. Herrmann, Physica A {\bf 191}, 263(1993); 
in {\em Proceedings of the III Granada School}, Lecture Notes in 
Physics Vol. 448 (Springer, Berlin, 1994).
\bibitem{FHP} U.~Frisch, B.~Hasslacher, and Y.~Pomeau, 
Phys. Rev. Lett. {\bf
56}, 1505(1986).
\bibitem{Kadanoff} L. P.~Kadanoff, G. R.~McNamara, and 
G.~Zanetti, Phys. Rev. 
E {\bf 40}, 4527(1989).
\bibitem{BHS} K.~Balasubramanian, F.~Hayot, and W. F.~Saam, Phys. Rev. 
E {\bf 36}, 2248(1987).
\bibitem{Rothman} D. H. Rothman and S. Zaleski, Rev. Mod. Phys. {\bf 66}, 
1417(1994).
\bibitem{Alonso} J. J.~Alonso and H. J.~Herrmann, Phys. Rev. Lett. 
 {\bf 76}, 4911(1996).
\bibitem{Kertesz} A.~K\'arolyi and J. Kert\'esz, in 
{\em Proceedings of the 6th Joint EPS--APS 
International Conference on Physics Computing} 
(European Physics Society, Geneva, Switzerland, 1994). 
\bibitem{Vollmar} S. Vollmar and H. J. Herrmann, Physica A {\bf 215}, 
411(1995).
\bibitem{Zanetti} I.~Goldhirsch and G.~Zanetti, Phys. Rev. Lett. {\bf
70}, 1619(1993); S. McNamara and W. R.~Young, Phys. Fluids A 
{\bf 5}, 34(1993).
\bibitem{Pouliquen} O.~Pouliquen and R.~Gutfraind, Phys. Rev. E 
{\bf 53}, 552(1996).
\bibitem{Ehrichs} E. E.~Ehrichs, H. M.~Jaeger, G. S.~Karczmar, 
J. B.~Knight, V. Y.~Kuperman, and S. R.~Nagel, Science {\bf 267}, 
1632(1995).
\end{thebibliography}
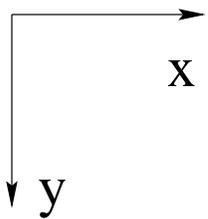
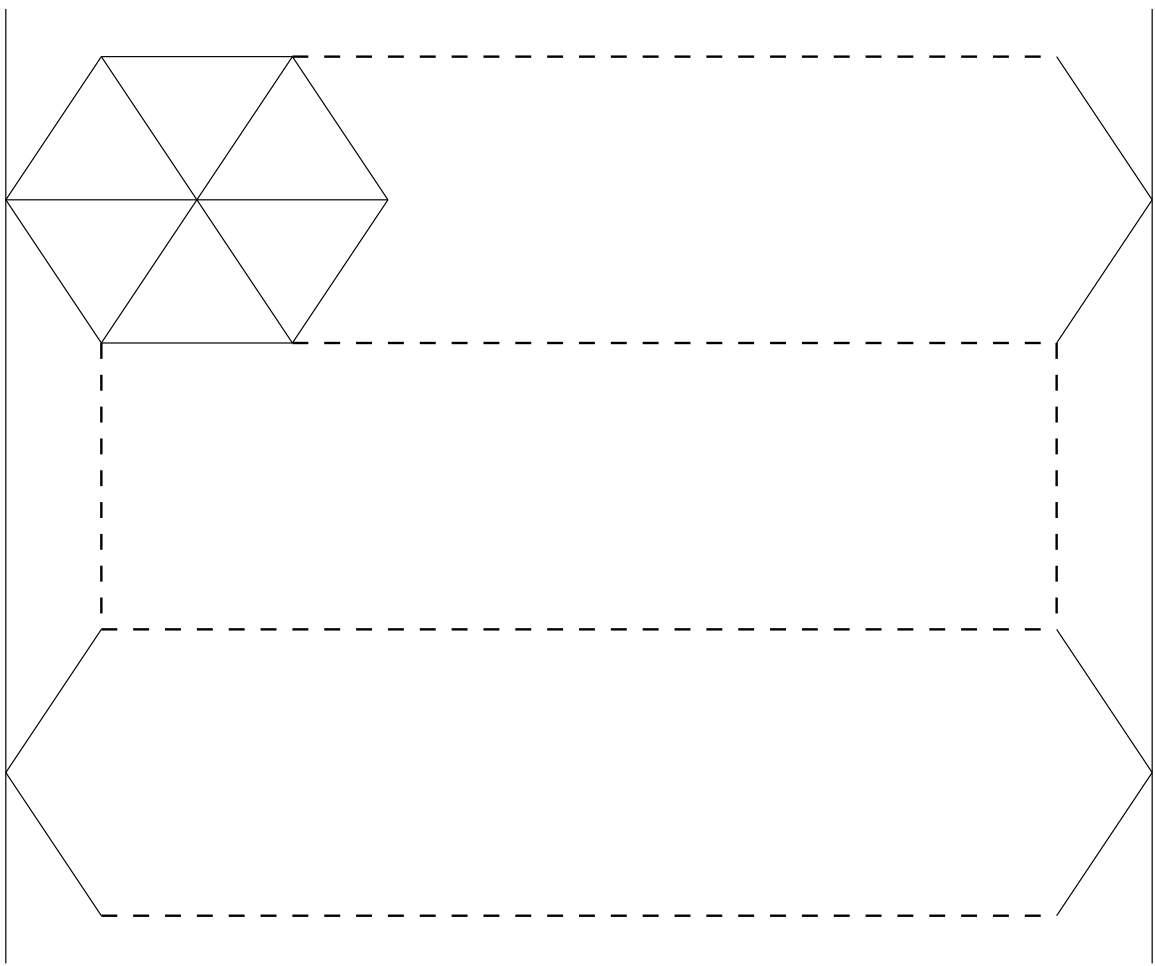
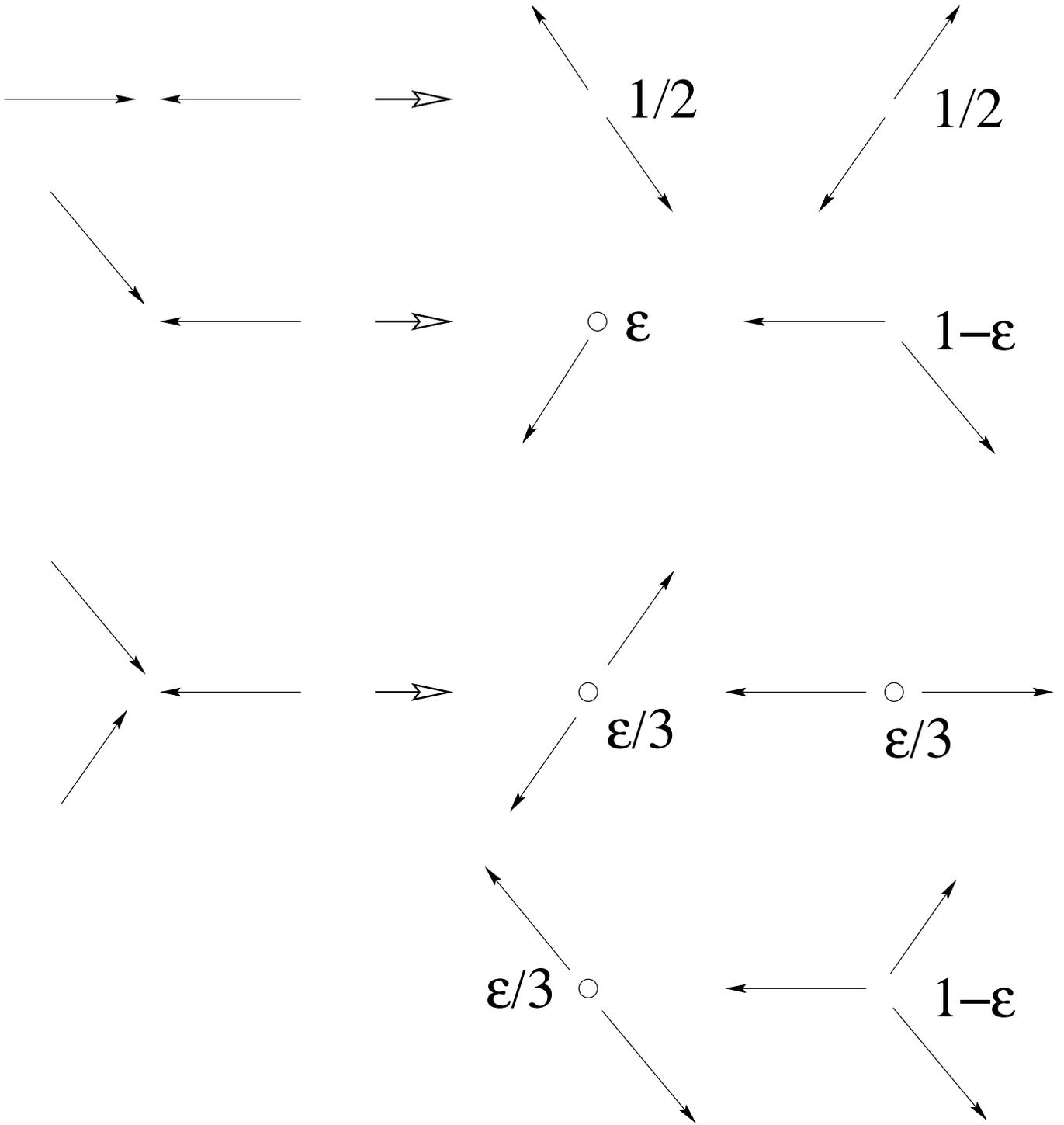
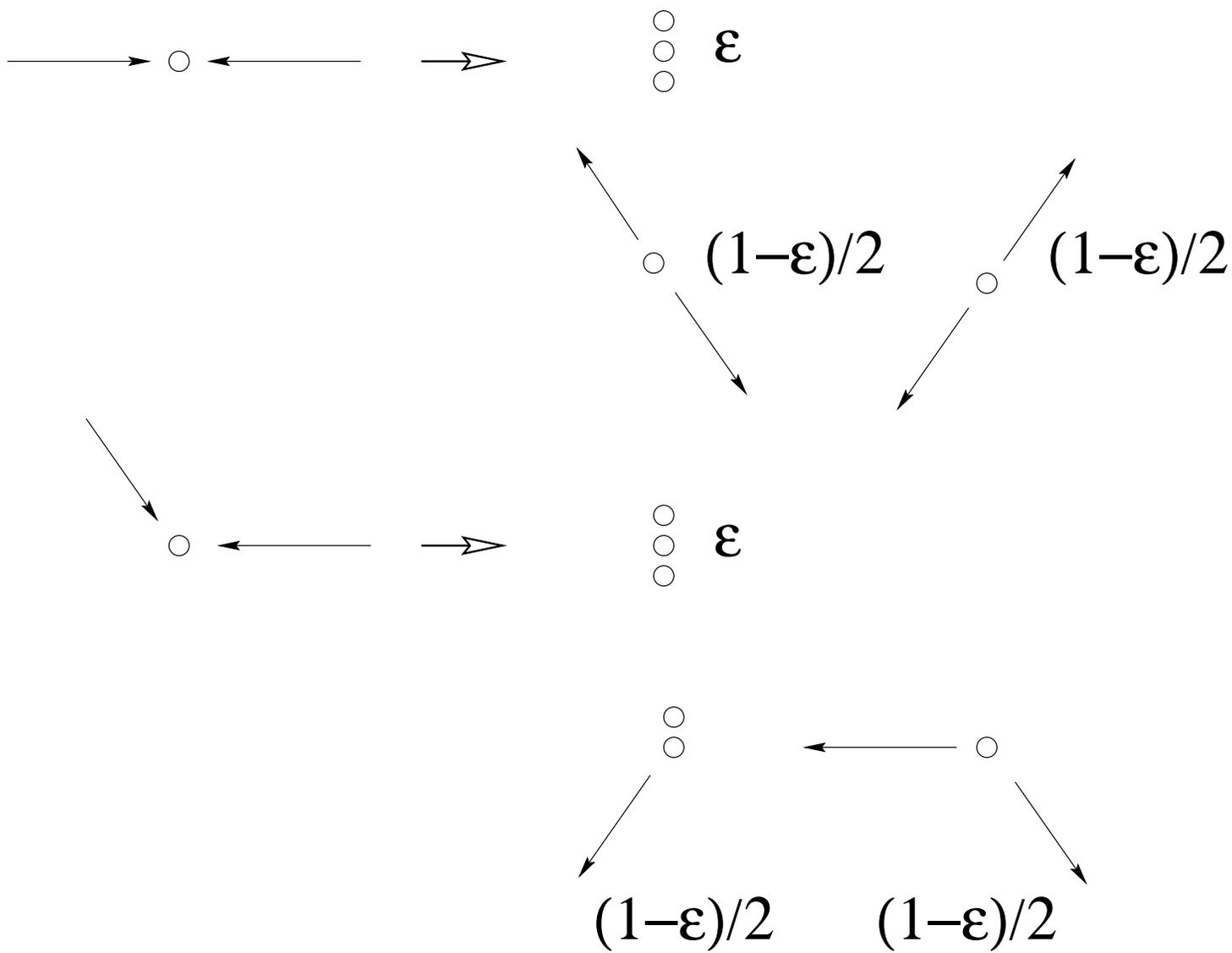
\end{document}